\documentclass[sigconf]{acmart}
\usepackage[linesnumbered, ruled, vlined]{algorithm2e}
\usepackage{xspace}
\usepackage{amsmath}
\usepackage{balance}
\usepackage{color}
\usepackage{graphicx, float, subfig, overpic}
\usepackage{setspace}
\usepackage{enumitem}
\usepackage{multirow, bm}
\usepackage{textcomp}

\DeclareMathOperator{\pr}{Pr}
\DeclareMathOperator{\var}{Var}

\SetCommentSty{myCommentStyle}

\renewcommand{\paragraph}[1]{
\vskip 1mm
\noindent
{\bf #1.}
}
\makeatletter
\def\subsubsection{\@startsection{subsubsection}{3}{\z@}%
    {-.2\baselineskip \@plus -2\p@ \@minus -.2\p@}%
    {.08\baselineskip}%
    {\ACM@NRadjust{\@subsubsecfont}}}
\makeatother

\theoremstyle{plain}
\newtheorem{theorem}{Theorem}[section]
\newtheorem{proposition}[theorem]{Proposition}
\newtheorem{lemma}[theorem]{Lemma}
\theoremstyle{definition}
\newtheorem{definition}[theorem]{Definition}

\DeclareMathSizes{9}{9}{6}{4}

\AtBeginDocument{%
    }

\setcopyright{acmlicensed}
\copyrightyear{2024}
\acmYear{2024}
\acmDOI{10.1145/3627673.3679759}

\acmConference[CIKM '24]{Proceedings of the 33rd ACM International Conference on Information and Knowledge Management}{October 21--25, 2024}{Boise, ID, USA}
\acmBooktitle{Proceedings of the 33rd ACM International Conference on Information and Knowledge Management (CIKM '24), October 21--25, 2024, Boise, ID, USA}
\acmISBN{979-8-4007-0436-9/24/10}

\begin{document}

\title{Privacy-Preserving Graph Embedding based on Local Differential Privacy}

\author{Zening Li}
\affiliation{%
    \institution{Beijing Institute of Technology}
    \city{Beijing}
    \country{China}
}
\email{zening-li@outlook.com}
\author{Rong-Hua Li}
\affiliation{%
    \institution{Beijing Institute of Technology}
    \city{Beijing}
    \country{China}
}
\email{lironghuabit@126.com}
\author{Meihao Liao}
\affiliation{%
    \institution{Beijing Institute of Technology}
    \city{Beijing}
    \country{China}
}
\email{mhliao@bit.edu.cn}
\author{Fusheng Jin}
\affiliation{%
    \institution{Beijing Institute of Technology}
    \city{Beijing}
    \country{China}
}
\email{jfs21cn@bit.edu.cn}
\author{Guoren Wang}
\affiliation{%
    \institution{Beijing Institute of Technology}
    \city{Beijing}
    \country{China}
}
\email{wanggrbit@gmail.com}


\begin{abstract}
Graph embedding has become a powerful tool for learning latent representations of nodes in a graph. Despite its superior performance in various graph-based machine learning tasks, serious privacy concerns arise when the graph data contains personal or sensitive information. To address this issue, we investigate and develop graph embedding algorithms that satisfy local differential privacy (LDP). We introduce a novel privacy-preserving graph embedding framework, named PrivGE, to protect node data privacy. Specifically, we propose an LDP mechanism to obfuscate node data and utilize personalized PageRank as the proximity measure to learn node representations. Furthermore, we provide a theoretical analysis of the privacy guarantees and utility offered by the PrivGE framework. Extensive experiments on several real-world graph datasets demonstrate that PrivGE achieves an optimal balance between privacy and utility, and significantly outperforms existing methods in node classification and link prediction tasks.
\end{abstract}

\ccsdesc[500]{Security and privacy~Privacy protections}
\ccsdesc[500]{Information systems~Data mining}

%
\keywords{Differential Privacy; Graph Embedding; Graph Neural Networks; Personalized PageRank}

\maketitle
\section{Introduction}\label{sec:introduction}
Many types of real-world data can be naturally represented as graphs, such as social networks, financial networks, and transportation networks. Over the past few years, graph embedding has attracted much attention due to its superior performance in various machine learning tasks, such as node classification and link prediction~\cite{yin2019scalable, qiu2019netsmf, zhang2021learning, wang2021approximate}. Graph embedding is a representation learning problem that uses vectors to capture node features and structural information in a graph. However, most real-world graphs involve sensitive information about individuals and their activities, such as users' profile information and comments in social networks. The direct release of embedding vectors of users provides a potential way for malicious attackers to infer the attributes and social interactions, which could potentially be private to the users. The growing awareness of privacy and the establishment of regulations and laws indicate that it is important to develop privacy-preserving graph embedding algorithms.

Differential privacy (DP)~\cite{dwork2006calibrating} has recently become the dominant paradigm for safeguarding individual privacy in data analysis. A vast majority of differentially private graph learning algorithms are designed under the centralized model~\cite{xu2018dpne, zhang2019graph, epasto2022differentially, daigavane2021node, kolluri2022lpgnet, sajadmanesh2023gap}, which assumes that a trusted data curator holds the personal data of all users and releases sanitized versions of statistics or machine learning models. However, this assumption is impractical in some applications due to security or logistical reasons. The centralized model carries the risk of users' data being breached by the trusted data curator through illegal access or internal fraud~\cite{imola2021locally, imola2021communication}. In addition, some social networks are inherently decentralized and distributed, where no centralized party holds the entire social graph. As a result, the centralized DP algorithms cannot be applied to these decentralized social networks.

In contrast to centralized DP, local differential privacy (LDP)~\cite{kasiviswanathan2011can} has emerged as a promising approach that ensures stronger privacy guarantees for users in scenarios involving data aggregation. LDP operates under a local model, where each user perturbs their data locally and transmits only the perturbed data to an untrusted data curator. This means that the original personal data remains confined to the users' local devices, effectively eliminating the risk of data leakage that might occur in the centralized DP settings. As a result, LDP provides an enhanced level of privacy protection for individuals.
The practicality and effectiveness of LDP have been recognized by major technology companies, such as Google~\cite{erlingsson2014rappor}, and Microsoft~\cite{ding2017collecting}, which have deployed LDP-based solutions to handle sensitive user data while preserving privacy. Moreover, the applicability of LDP extends beyond centralized settings, making it an appealing choice for decentralized applications. 

In this paper, our focus is on exploring the design of graph embedding algorithms that satisfy LDP, where the node features are private and sensitive, and the global graph structure information is maintained by the data curator. This scenario arises in various domains, particularly in social network analysis and mobile computing. For instance, consider a social networking platform or a dating application that utilizes node representations to capture relationships between users. In this context, individuals' attributes and preferences are treated as private node features. Integrating LDP in graph embedding for such applications can safeguard users' sensitive data while empowering the platform to offer valuable services.

In the local setting for DP, each user sends only perturbed node features to the data collector. However, a key challenge arises when dealing with high-dimensional node features used for graph embedding, as the perturbation in such scenarios can result in significant information loss. To overcome this challenge, some researchers have proposed various approaches, including sampling techniques and tailored perturbation mechanisms, aimed at preserving utility in the high-dimensional space~\cite{duchi2018minimax, wang2019collecting, sajadmanesh2021locally, jin2022gromov, lin2022towards}. Nevertheless, these mechanisms also introduce excessive noise to the data, potentially compromising overall performance.

In this paper, we propose PrivGE, a novel privacy-preserving graph embedding framework based on private node data. Our framework offers provable privacy guarantees, building on the principles of local differential privacy. Specifically, to protect the privacy of node features, we propose the HDS (an acronym for \underline{H}igh-\underline{D}imensional \underline{S}quare wave) mechanism, an LDP perturbation technique tailored for high-dimensional data. Each user can adopt this perturbation mechanism to obfuscate their features before sending them to the data curator. The server leverages graph structure information and perturbed node features to learn graph representations. To avoid neighborhood explosion and over-smoothing issues, we decouple the feature transformation from the graph propagation. Furthermore, we adopt personalized PageRank as the proximity measure to learn node representations. Importantly, we conduct a comprehensive theoretical analysis of the utility of the PrivGE framework. Our findings indicate that the proposed approach yields smaller error bounds than existing mechanisms, specifically, from $\mathcal{O}(\frac{d\log(d/\delta)}{\epsilon})$ down to $\mathcal{O}(\log(d/\delta)$), making it a more efficient solution\footnote[1]{Note that $d$ denotes the dimension of the node features, and $\epsilon$ represents the privacy budget. Additionally, $\delta$ is a constant between $(0, 1]$.}. Finally, to assess the effectiveness of the PrivGE framework, we conduct extensive experiments on various real-world datasets. The results demonstrate that our proposed method establishes state-of-the-art performance and achieves decent privacy-utility trade-offs in node classification and link prediction tasks. In summary, we highlight the main contributions as follows:
\begin{itemize}[leftmargin=*]
\item We propose PrivGE, an innovative framework that aims to preserve privacy in graph embedding. Our method provides provable privacy guarantees and simultaneously ensures effective graph representation learning.
\item To address the challenge dealing with high-dimensional node features, we propose the HDS mechanism to protect node feature privacy. This perturbation technique empowers users to obfuscate their features locally before reporting them to the data curator, thus enhancing privacy protection.
\item We conduct a comprehensive theoretical analysis of the utility of PrivGE and alternative mechanisms. The results demonstrate that our mechanism offers smaller error bounds than the others, reducing them from $\mathcal{O}(\frac{d\log(d/\delta)}{\epsilon})$ to $\mathcal{O}(\log(d/\delta)$).
\item We conduct extensive experiments on various real datasets. The experimental results show that our proposed method achieves better privacy-utility trade-offs than existing solutions. For instance, our proposed method achieves about $ 8\% $ higher accuracy than the best competitor on the Pubmed dataset in node classification.
\end{itemize}

\section{Preliminaries} \label{sec:preliminaries}

\paragraph{Problem Statement}
We consider an undirected and unweighted graph $G=(\mathcal{V}, \mathcal{E})$, where $\mathcal{V}$ is the set of nodes (i.e., users) and $\mathcal{E}$ represents the set of edges. Let $|\mathcal{V}|$ be the number of nodes. Each user $v \in \mathcal{V}$ is characterized by a $d$-dimensional feature vector $\mathbf{x}_v$, and we use $\mathbf{X} \in \mathbb{R}^{|\mathcal{V}| \times d}$ to denote the feature matrix. Without loss of generality, we assume the node features are normalized into $[-1, 1]$\footnote[2]{Note that it is a common assumption in~\cite{sajadmanesh2021locally} that the feature fields are known to the users, so this normalization step does not compromise privacy.}. Let $\mathbf{A}$ and $\mathbf{D}$ represent the adjacency matrix and the diagonal degree matrix, respectively. For each node $v \in \mathcal{V}$, $\mathcal{N}(v)$ is the set of neighbors of $v$, and the degree of $v$ is $|\mathcal{N}(v)|$.

We assume that the data curator is an untrusted party with access to the node set $\mathcal{V}$ and edge set $\mathcal{E}$. However, the data curator cannot observe the feature matrix $\mathbf{X}$, which is private to users. Our ultimate objective is to learn a node embedding matrix and simultaneously protect the privacy of node data.

\paragraph{Local Differential Privacy}
Differential privacy~\cite{dwork2006calibrating} has become the dominant model for the protection of individual privacy from powerful and realistic adversaries. DP can be bifurcated into centralized DP and local DP. Centralized DP assumes a scenario where a trusted curator holds all users' personal data and releases sanitized versions of the statistics. In contrast to centralized DP, local DP operates under the assumption of a local model, where the data curator is considered untrusted. Specifically, each user perturbs their data via a randomized perturbation mechanism and sends the obfuscated data to the untrusted data curator. 

\begin{definition}[$\epsilon$-Local Differential Privacy~\cite{kasiviswanathan2011can}] \label{def:LDP}
    Given $\epsilon > 0$, a randomized algorithm $\mathcal{A}$ satisfies $\epsilon$-local differential privacy if and only if for any two users' private data $x$ and $x^{\prime}$, and for any possible output $y \in Range(\mathcal{A})$, we have
    \begingroup
    \setlength{\abovedisplayskip}{3pt}
    \setlength{\belowdisplayskip}{3pt}
    \begin{align}
        \pr[\mathcal{A}(x)=y] \leq e^{\epsilon} \cdot \pr[\mathcal{A}(x^{\prime})=y].
    \end{align}
    \endgroup
\end{definition}
Here, the parameter $\epsilon$ is called the privacy budget, which controls the strength of privacy protection: a lower privacy budget indicates stronger privacy preservation but leads to lower utility. In addition, LDP satisfies some important properties that can help us develop more sophisticated algorithms.
\begin{proposition}[Sequential Composition~\cite{day2016publishing}] \label{sequential-composition}
    Given the sequence of computations $\mathcal{A}_{1}, \mathcal{A}_{2}, \dots, \mathcal{A}_{k}$, if each $\mathcal{A}_{i}$ satisfies $\epsilon_{i}$-LDP, then their sequential execution on the same dataset satisfies $\sum_{i}\epsilon_{i}$-LDP.
\end{proposition}
\begin{proposition}[Post-processing~\cite{day2016publishing}] \label{post-processing}
    Given $\mathcal{A}(\cdot)$ that satisfies $\epsilon$-LDP, then for any algorithm $\mathcal{B}$, the composed algorithm $\mathcal{B}(\mathcal{A}(\cdot))$ also satisfies $\epsilon$-LDP.
\end{proposition}

\paragraph{Graph Embedding}
The task of graph embedding is to learn the latent representation of each node. Numerous studies have shown that the latent representations can capture the structural and inherent properties of the graph, which can facilitate downstream inference tasks, such as node classification and link prediction~\cite{yin2019scalable, qiu2019netsmf, zhang2021learning}.

As an important class of graph embedding methods, the message passing framework is of interest due to its flexibility and favorable performance. This framework comprises two phases: (i) message propagation among neighbors, and (ii) message aggregation to update representation. Most GNN models, such as GCN~\cite{kipf2016semi} and GAT~\cite{velivckovic2017graph}, employ a message passing process to spread information. At each layer, feature transformation is coupled with aggregation and propagation. Increasing the number of layers allows the model to incorporate information from more distant neighbors, which promotes a more comprehensive node representation. However, this approach may lead to over-smoothing and neighborhood explosion~\cite{bojchevski2020scaling}.

To address these inadequacies, some studies~\cite{klicpera2018predict, bojchevski2020scaling} decouple the feature transformation from the graph propagation and exploit node proximity queries to incorporate multi-hop neighborhood information. Personalized PageRank~\cite{bojchevski2020scaling}, a widely-used proximity measure, can characterize node distances and similarities. Consequently, we apply the personalized PageRank matrix to the feature matrix $\mathbf{X}$ to derive the representation matrix $\mathbf{Z}$. The graph propagation equation is defined as follows:
\begingroup
\setlength{\abovedisplayskip}{3pt}
\setlength{\belowdisplayskip}{3pt}
\begin{align} \label{eq:graph-propagation}
    \mathbf{Z} = \mathbf{\Pi} \cdot \mathbf{X} = \sum\limits_{\ell=0}^{\infty} \alpha (1-\alpha)^{\ell} \cdot (\mathbf{D}^{r-1}\mathbf{A}\mathbf{D}^{-r})^{\ell} \cdot \mathbf{X},
\end{align}
\endgroup
where $\mathbf{\Pi}$ is the personalized PageRank matrix, $r \in [0, 1]$ is the convolution coefficient and $\alpha \in (0, 1)$ is the decay factor. The parameter $\alpha$ controls the amount of information we capture from the neighborhood.
To be specific, for the values of $\alpha$ closer to $1$, we place more emphasis on the immediate neighborhood of the node, which can avoid over-smoothing. As the value of $\alpha$ decreases to $0$, we instead give more attention to the multi-hop neighborhood of the node.

\section{The Proposed Method} \label{sec:proposed-method}
In this section, we describe our proposed differentially private framework for graph embedding. It consists of two components: a perturbation module and a propagation module. The perturbation module is utilized to locally obfuscate node features before they are sent to the data curator. This component helps relieve users' concerns about sharing their private information. 
However, the node features to be collected are likely high-dimensional, whereas most LDP perturbation functions focus on one-dimensional data, such as the Laplace mechanism. Since each user is authorized a limited privacy budget, the allocated privacy budget in each dimension is diluted as the number of dimensions increases, which results in more noise injection. While some LDP mechanisms, such as One-bit mechanism~\cite{ding2017collecting} and Piecewise mechanism~\cite{wang2019collecting}, have been extended to handle the high-dimensional data~\cite{sajadmanesh2021locally, wang2019collecting}, these mechanisms introduce much noise into the data, which compromises performance. The Square wave mechanism~\cite{li2020estimating} is another LDP mechanism that aims to reconstruct the distribution of one-dimensional numerical attributes. This mechanism provides a more concentrated perturbation than the two mechanisms mentioned above (i.e., One-bit mechanism and Piecewise mechanism). Thus, to address this problem, we extend the Square wave mechanism to handle high-dimensional data, and develop an LDP mechanism to perturb the node features.

The propagation module is used to spread node features via information exchange between adjacent nodes, where the representation of each node is updated based on the aggregation of its neighbors' features. However, most methods suffer from neighborhood explosion and over-smoothing issues, as explained in Section~\ref{sec:preliminaries}. To address these problems, we decouple feature transformation and propagation, and adopt personalized PageRank as the graph propagation formula to obtain the representation matrix. More importantly, we provide a comprehensive theoretical analysis of the utility of our proposed method and alternative mechanisms.

In the rest of this section, we first introduce the technical details of the perturbation module used for privacy assurances and some theoretical properties of our proposed mechanism. Next, we present the propagation process designed for graph embedding. Finally, we conduct a utility analysis of the proposed framework.

\subsection{Perturbation Module} \label{subsec:perturbation-module}
The target of the perturbation module is to gather node features from individuals under LDP. In specific, each user $v \in \mathcal{V}$ perturbs his/her private feature vector $\mathbf{x}_v$ using the perturbation mechanism and sends the perturbed data $\tilde{\mathbf{x}}_v$ to the data curator. The crucial aspect is to devise a randomization mechanism that provides plausible deniability. In this section, we first review three existing perturbation mechanisms and discuss their deficiencies. Then, we propose our mechanism for feature perturbation.

\paragraph{Existing Solutions}
Laplace mechanism~\cite{dwork2006calibrating} is a well-established approach enforcing differential privacy. It can be applied to the LDP setting in the following manner. Assuming each user possesses a one-dimensional value $x$ in the range of $[-1, 1]$, we define a randomized function that generates a perturbed value $\tilde{x} = x + Lap(2/\epsilon)$. Here, $Lap(\lambda)$ represents a random variable that follows a Laplace distribution with a scale parameter $\lambda$. The Laplace distribution is characterized by the probability density function $f(x) = \frac{1}{2\lambda}\exp(-\frac{|x|}{\lambda})$.

To extend this mechanism to high-dimensional values, a straightforward method is to collect the perturbed value separately using the Laplace mechanism in each dimension. In this approach, each dimension is assigned a privacy budget of $\epsilon / d$. Applying the composition property of LDP as described in Proposition~\ref{sequential-composition}, we can conclude that the entire collection of values satisfies $\epsilon$-LDP. In the high-dimensional space, since the injected Laplace noise in each dimension follows $Lap(2d/\epsilon)$, it is evident that the perturbed value $\tilde{x}$ is unbiased, and its variance is $\frac{8d^2}{\epsilon^2}$. Furthermore, the Laplace mechanism embodies a category of LDP mechanisms known as unbounded mechanisms, where the noise injected into the original value ranges from negative to positive infinity.

In the one-dimensional Piecewise mechanism~\cite{wang2019collecting}, the input domain is $[-1, 1]$, and the range of perturbed data is $[-s, s]$, where $s = \frac{e^{\epsilon / 2} + 1}{e^{\epsilon / 2} - 1}$. Given an original value $x \in [-1, 1]$, the perturbed value $\tilde{x}$ is sampled from the following distribution:
\begingroup
\setlength{\abovedisplayskip}{3pt}
\setlength{\belowdisplayskip}{3pt}
\begin{align}
    \pr[\tilde{x}=c|x] = 
    \begin{cases}
        p, & \text{if} \ c \in [\ell(x), r(x)], \\[-0.2em]
        \frac{p}{e^{\epsilon}}, & \text{if} \ c \in [-s, \ell(x)) \cup (r(x), s],
    \end{cases}
\end{align}
\endgroup
where $ p = \frac { e^{\epsilon} - e^{\epsilon / 2}}{2e^{\epsilon / 2} + 2}$, $\ell(x)=\frac{s+1}{2} \cdot x - \frac{s-1}{2}$, and $ r(x) = \ell(x)+s-1 $.
Wang et al.~\cite{wang2019collecting} also propose an extension of the Piecewise mechanism to process high-dimensional data. The extended mechanism adopts a sampling technique so that each user reports only $k$ out of $d$ dimensions of his/her perturbed data to the data curator. In that case, each reporting dimension is allocated $\epsilon/ k$ privacy budget, and the reporting data $\hat{x}$ is calibrated to ensure that the final outcome $\tilde{x}$ is unbiased. Formally, $\tilde{x}$ is obtained by $ \tilde{x} = \frac{d}{k} \cdot \hat{x} $.
In the high-dimensional setting, the variance of $\tilde{x}$ induced by Piecewise mechanism is $ \var[\tilde{x}] = \frac{d(e^{\epsilon/ (2k)}+3)}{3k(e^{\epsilon/(2k)}-1)^{2}} + [\frac{d\cdot e^{\epsilon / (2k)}}{k(e^{\epsilon/(2k)}-1)}-1]\cdot x^2 $.

Multi-bit mechanism~\cite{sajadmanesh2021locally} is another perturbation function used to handle high-dimensional data under LDP. In its one-dimensional form, the original data $x \in [-1, 1]$ is perturbed into $-1$ or $1$, with the following probabilities:
\begingroup
\setlength{\abovedisplayskip}{3pt}
\setlength{\belowdisplayskip}{3pt}
\begin{align}
    \pr[\tilde{x}=c|x] =
    \begin{cases}
        \frac{1}{e^{\epsilon} + 1} + \frac{x+1}{2} \cdot \frac{e^{\epsilon} - 1}{e^{\epsilon} + 1}, & \text{if} \ c = 1, \\[-0.2em]
        \frac{e^{\epsilon}}{e^{\epsilon} + 1} - \frac{x+1}{2} \cdot \frac{e^{\epsilon} - 1}{e^{\epsilon} + 1}, & \text{if} \ c  = -1.
    \end{cases}\
\end{align}
\endgroup

\begin{figure}[!t]
    \centering
    \includegraphics[width=0.65\columnwidth]{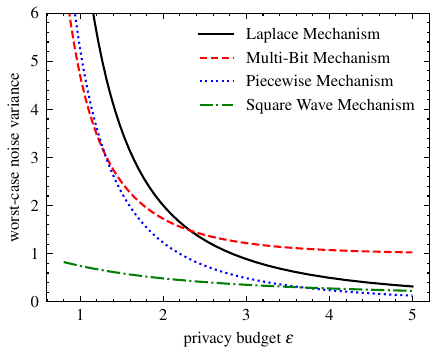}
    \caption{The worst-case noise variance vs. privacy budget for one-dimensional data.}
    \label{fig:worse-case-noise-variance}
    \Description{worse case noise variance}
    \vspace*{-0.1cm}
\end{figure}

In the high-dimensional space, similar to the Piecewise mechanism, the algorithm first uniformly samples $k$ out of $d$ dimensions without replacement and then performs $\frac{\epsilon}{k}$-LDP perturbation for each sampled dimension. In the end, the data curator transforms the reporting data $\hat{x}$ to its unbiased estimate $ \tilde{x} = \frac{d(e^{\epsilon / k } + 1)}{k(e^{\epsilon / k } - 1)} \cdot \hat{x} $.
The variance of $\tilde{x}$ induced by Multi-bit mechanism is $ \var[\tilde{x}] = \frac{d}{k}(\frac{e^{\epsilon / k}+1}{e^{\epsilon / k}-1})^{2} - x^2 $.
In contrast to the Laplace mechanism, the Piecewise and Multi-bit mechanisms perturb the original value into a bounded domain. Consequently, these mechanisms are referred to as bounded mechanisms.

\paragraph{Deficiencies of Existing Solutions}
Even though these mechanisms can handle high-dimensional data, they introduce a significant amount of noise to the private data, leading to a decline in performance. To be specific, in the one-dimensional scenario, we visualize the worst-case noise variance of different mechanisms under varying privacy budgets, as illustrated in Figure~\ref{fig:worse-case-noise-variance}. Note that in Figure~\ref{fig:worse-case-noise-variance}, we also add the Square wave mechanism~\cite{li2020estimating} for comparison. The Square wave mechanism was originally proposed in~\cite{li2020estimating}, which can achieve LDP when handling one-dimensional data. We can observe that the Square wave mechanism~\cite{li2020estimating} provides a considerably smaller noise variance than the Piecewise mechanism for $\epsilon \leq 3.5$, and only slightly larger than the latter for $\epsilon > 3.5$. Moreover, the worst-case noise variance provided by the Square wave mechanism is consistently smaller than that of the Laplace mechanism and the Multi-bit mechanism when $\epsilon \leq 5.0$. In consequence, the Square wave mechanism provides more concentrated perturbation. In the high-dimensional setting, unbiased calibration is not conducive to graph embedding and can result in injecting excessive noise. The Square wave mechanism is designed to estimate the distribution of one-dimensional numerical data. Inspired by the ideas of~\cite{wang2019collecting} and~\cite{sajadmanesh2021locally}, we generalize the Square wave mechanism to feature collection in high-dimensional space.

\begin{algorithm}[!t]
    \setstretch{0.9}
    \small
    \caption{Extended Square Wave Mechanism}
    \label{alg:ods}
    \KwIn{single feature $x \in [-1, 1]$, privacy budget $\epsilon > 0$}
    \KwOut{perturbed feature $\tilde{x} \in [-b-1, 1+b]$}
    Let $b = \frac{\epsilon e^{\epsilon} - e^{\epsilon} + 1}{e^{\epsilon}(e^{\epsilon} - \epsilon - 1)}$\;
    Let $\eta $ be sampled uniformly from $[0, 1]$\;
    \If{$\eta < \frac{be^{\epsilon}}{be^{\epsilon} + 1}$}{
        Sample a random value $\tilde{x}$ uniformly from $[x-b, x+b]$\;
    }
    \Else{
        Sample a random value $\tilde{x}$ uniformly from $[-b-1, x-b)\cup (x+b, 1+b]$\;
    }
    \Return{$\tilde{x}$}
\end{algorithm}

\paragraph{HDS Mechanism}
Our HDS mechanism is built upon the Square wave mechanism~\cite{li2020estimating}, which can handle only one-dimensional data. The Square wave mechanism is based on the following intuition. Given a single feature $x$, the perturbation module should report a value close to $x$ with a higher probability than a value far away from $x$. To some extent, the value close to $x$ also carries useful information about the input. The Square wave mechanism is initially designed for an input domain of $[0, 1]$. However, in our scenario, the node features are normalized to the range of $[-1, 1]$. Consequently, we extend this mechanism to enhance its capability in handling a broader range of node features, specifically $[-1, 1]$. Algorithm~\ref{alg:ods} outlines the perturbation process for one-dimensional data. The algorithm takes a single feature $x \in [-1, 1]$ as input and produces a perturbed feature $\tilde{x} \in [-b-1, 1+b]$, where $b = \frac{\epsilon e^{\epsilon} - e^{\epsilon} + 1}{e^{\epsilon}(e^{\epsilon} - \epsilon - 1)}.$ Let $p = \frac {e^{\epsilon}}{2be^{\epsilon} + 2}$ and $q = \frac{1}{2be^{\epsilon} + 2}$. The noisy output $\tilde{x}$ follows the distribution as below:
\begingroup
\setlength{\abovedisplayskip}{3pt}
\setlength{\belowdisplayskip}{3pt}
\begin{align} \label{eq:ods}
    \pr[\tilde{x}=c|x] =
    \begin{cases}
        p, &\text{if } c \in [x\!-\!b, x\!+\!b], \\[-0.2em]
        q, &\text{if } c \in [-b\!-\!1, x\!-\!b)\! \cup\! (x\!+\!b, 1\!+\!b].
    \end{cases}
\end{align}
\endgroup

Algorithm~\ref{alg:hds} presents the pseudo-code of our HDS mechanism for high-dimensional feature collection. The algorithm requires that each user perturbs only $k$ dimensions of their features vector rather than $d$. This is because, according to the composition property of LDP described in Proposition~\ref{sequential-composition}, it increases the privacy budget for each dimension from $\epsilon / d$ to $\epsilon / k$, reducing the noise variance. Given a feature vector $\mathbf{x}$, the algorithm first uniformly samples $k$ values from $\{1, 2, \dots, d\}$ without replacement, where $k$ is a parameter that controls the number of dimensions to be perturbed. Then for each sampled value $j$, the perturbed feature $\tilde{x}_j$ is generated by Algorithm~\ref{alg:ods}, taking $x_j$ and $\frac{\epsilon}{k}$ as input. Correspondingly, the rest of the $d-k$ features are encoded into $0$ to prevent privacy leakage. Given that the output domain of the HDS mechanism is bounded, our proposed mechanism can be categorized as a bounded mechanism.

\begin{algorithm}[!t]
    \setstretch{0.9}
    \small
    \caption{HDS Mechanism}
    \label{alg:hds}
    \KwIn{feature vector $\mathbf{x} \in [-1, 1]^{d}$, privacy budget $\epsilon > 0$, sampling parameter $k \in \{1,2,...,d\}$}
    \KwOut{perturbed feature vector $\tilde{\mathbf{x}} \in [-b-1, 1+b]^{d}$}
    Let $\mathcal{S}$ be a set of $k$ values sampled uniformly without replacement from $\{1, 2, \dots, d\}$\;
    \For{$j \in \{1, 2, \dots, d\}$}{
        \If{$j \in \mathcal{S}$}{
            $\tilde{x}_{j} \gets $ Feed $x_{j}$ and $\frac{\epsilon}{k}$ as input to Algorithm~\ref{alg:ods}\;
        }
        \Else{
            $\tilde{x}_{j} \gets 0$\;
        }
    }
    \Return{$\tilde{\mathbf{x}} = [\tilde{x}_{1}, \tilde{x}_{2}, \dots, \tilde{x}_{d}]^{\top}$}
\end{algorithm}

\begin{theorem}
    The HDS mechanism presented in Algorithm~\ref{alg:hds} satisfies $\epsilon$-local differential privacy for each node.
\end{theorem}
\begin{proof}
    First, we prove that the Algorithm~\ref{alg:ods} provides $\epsilon$-LDP. Let $\mathcal{A}(x)$ be the Algorithm~\ref{alg:ods} that takes the single feature $x$ as input, and $\tilde{x} = \mathcal{A}(x)$ is the obfuscated feature corresponding to $x$. Suppose $x_{1}$ and $x_{2}$ are private features of any two users. According to Equation~(\ref{eq:ods}), for any output $\tilde{x} \in [-b-1, 1+b]$, we have $ \frac{\pr[\mathcal{A}(x_{1}) = \tilde{x}]}{\pr[\mathcal{A}(x_{2}) = \tilde{x}]} \leq \frac{e^{\epsilon}}{2be^{\epsilon} + 2}\cdot(2be^{\epsilon} + 2)=e^{\epsilon} $.
    Thus, the Algorithm~\ref{alg:ods} satisfies $\epsilon$-LDP. Since Algorithm~\ref{alg:hds} executes $\frac{\epsilon}{k}$-LDP operations (Algorithm~\ref{alg:ods}) $k$ times on the same input data, then according to the composition property, Algorithm~\ref{alg:hds} satisfies $\epsilon$-local differential privacy.
\end{proof}

In the following analysis, we examine the bias and variance of the HDS mechanism.
\begin{lemma} \label{lemma:ohs-bias-variance}
    Let $\tilde{\mathbf{x}}_{v}$ be the output of Algorithm~\ref{alg:hds} on the input vector $\mathbf{x}_v$. For any dimension $j \in \{1, 2, \dots, d\}$, $ \mathbb{E}[\tilde{x}_{v, j}] = C \cdot x_{v, j} $ and $ \var[\tilde{x}_{v, j}] = \frac{k(b^{3}e^{\epsilon / k} + 3b^{2} + 3b + 1)}{3d(be^{\epsilon / k}+1)} + (C-C^{2})x_{v, j}^{2} $, 
    where $C = \frac{kb(e^{\epsilon / k} - 1)}{dbe^{\epsilon / k}+1}$.
\end{lemma}
\begin{proof}
    For the expectation, we have
    \begingroup
    \setlength{\abovedisplayskip}{3pt}
    \setlength{\belowdisplayskip}{3pt}
    \begin{align} \label{eq:hds-bias-1}
        & \mathbb{E}[\tilde{x}_{v, j}] = \mathbb{E}[\tilde{x}_{v, j}|j\in \mathcal{S}]\pr[j\in \mathcal{S}] + \mathbb{E}[\tilde{x}_{v, j}|j \notin \mathcal{S}]\pr[j \notin \mathcal{S}] \nonumber \\[-0.2em]
        &\! = \frac{k}{d} \cdot \mathbb{E}[\tilde{x}_{v, j}|j\in \mathcal{S}].
    \end{align}
    \endgroup
    According to Equation~(\ref{eq:ods}), we have
    \begingroup
    \setlength{\abovedisplayskip}{3pt}
    \setlength{\belowdisplayskip}{3pt}
    \begin{align} \label{eq:ods-bias}
        & \mathbb{E}[\tilde{x}_{v, j}|j\in \mathcal{S}]\! =\! \frac{1}{2be^{\epsilon / k}\! +\! 2}\left(\int_{-1-b}^{x_{v, j}-b} tdt\!+\!\int_{x_{v, j}-b}^{x_{v, j}+b} te^{\epsilon / k}dt\!+\!\int_{x_{v, j}+b}^{1+b} tdt\right) \nonumber \\[-0.2em]
        &\! = \frac{b(e^{\epsilon / k} - 1)}{be^{\epsilon / k}+1} \cdot x_{v, j}.
    \end{align}
    \endgroup
    Combining (\ref{eq:hds-bias-1}) and (\ref{eq:ods-bias}) we conclude
    \begingroup
    \setlength{\abovedisplayskip}{3pt}
    \setlength{\belowdisplayskip}{3pt}
    \begin{align} \label{eq:hds-bias}
        \mathbb{E}[\tilde{x}_{v, j}] = \frac{kb(e^{\epsilon / k} - 1)}{dbe^{\epsilon / k}+1} \cdot x_{v, j} = C \cdot x_{v, j}.
    \end{align}
    \endgroup
    For the variance, we have
    \begingroup
    \setlength{\abovedisplayskip}{3pt}
    \setlength{\belowdisplayskip}{3pt}
    \begin{align} \label{eq:hds-variance-1}
        & \var[\tilde{x}_{v, j}] = \mathbb{E}[\tilde{x}_{v, j}^{2}] - (\mathbb{E}[\tilde{x}_{v, j}])^{2} \nonumber \\[-0.2em]
        & \! = \mathbb{E}[\tilde{x}_{v, j}^{2}|j\in \mathcal{S}]\pr[j\in \mathcal{S}]\! +\! \mathbb{E}[\tilde{x}_{v, j}^{2}|j \notin \mathcal{S}]\pr[j \notin \mathcal{S}]\! -\! \mathbb{E}^{2}[\tilde{x}_{v, j}] \nonumber \\[-0.2em]
        & \! = \frac{k}{d} \cdot \mathbb{E}[\tilde{x}_{v, j}^{2}|j\in \mathcal{S}] - (\mathbb{E}[\tilde{x}_{v, j}])^{2}.
    \end{align}
    \endgroup
    According to Equation~(\ref{eq:ods}), we have
    \begingroup
    \setlength{\abovedisplayskip}{3pt}
    \setlength{\belowdisplayskip}{3pt}
    \begin{align} \label{eq:ods-variance}
        & \mathbb{E}[\tilde{x}_{v, j}^2|j\in \mathcal{S}]\! =\! \frac{\int_{-1-b}^{x_{v, j}-b} t^2 \,dt\! +\! \int_{x_{v, j}-b}^{x_{v, j}+b} t^{2}e^{\epsilon / k} \,dt\! +\! \int_{x_{v, j}+b}^{1+b} t^{2} \,dt}{2be^{\epsilon / k} + 2} \nonumber \\[-0.2em]
        & \! = \frac{b^{3}e^{\epsilon / k} + 3b^{2} + 3b + 1}{3(be^{\epsilon / k}+1)} + \frac{b(e^{\epsilon / k} - 1)}{be^{\epsilon / k}+1} \cdot x_{v, j}^{2}.
    \end{align}
    \endgroup
    Combining (\ref{eq:hds-bias}), (\ref{eq:hds-variance-1}) and (\ref{eq:ods-variance}) yields
    \begingroup
    \setlength{\abovedisplayskip}{3pt}
    \setlength{\belowdisplayskip}{3pt}
    \begin{align} \label{eq:hds-variance}
        \var[\tilde{x}_{v, j}] = \frac{k(b^{3}e^{\epsilon / k} + 3b^{2} + 3b + 1)}{3d(be^{\epsilon / k}+1)}\! +\! (C-C^{2})x_{v, j}^{2}.
    \end{align}
    \endgroup
\end{proof}

\subsection{Propagation Module} \label{subsec:propagation-module}
\begin{algorithm}[t!]
    \setstretch{0.9}
    \small
    \caption{Backward Push Propagation}
    \label{alg:graph-propagation}
    \KwIn{graph $G=(\mathcal{V}, \mathcal{E})$, perturbed feature matrix $\tilde{\mathbf{X}}$, decay factor $\alpha$, convolutional coefficient $r$, threshold $r_{max}$}
    \KwOut{embedding matrix $\tilde{\mathbf{Z}}$}
    Initialize reserve matrix $\mathbf{Q} \gets \mathbf{0}$ and residue matrix $\mathbf{R} \gets \mathbf{D}^{-r}\tilde{\mathbf{X}}$\;
    \While{$\exists v$ and $\exists j \in \{0, \dots, d-1\}$ s.t. $|\mathbf{R}(v, j)| > r_{max}$}{
        \For{$u \in \mathcal{N}(v)$}{
            $\mathbf{R}(u, j) \gets \mathbf{R}(u, j) + (1-\alpha) \cdot \frac{\mathbf{R}(v, j)}{|\mathcal{N}(u)|}$\;
        }
        $\mathbf{Q}(v, j) \gets \mathbf{Q}(v, j) + \alpha \cdot \mathbf{R}(v, j)$\;
        $\mathbf{R}(v, j) \gets 0$\;
    }
    $\tilde{\mathbf{Z}} \gets \mathbf{D}^{r} \cdot \mathbf{Q}$\;
    \Return{$\tilde{\mathbf{Z}}$}
\end{algorithm}
The propagation module takes the perturbed feature matrix $\tilde{\mathbf{X}}$ as input, which comprises obfuscated feature vectors $\tilde{\mathbf{x}}_v$ for each user $v \in \mathcal{V}$, and outputs the embedding matrix $\tilde{\mathbf{Z}}$. To incorporate neighborhood information, we exploit personalized PageRank as the proximity measure for graph propagation. Formally, the process of graph propagation can be formulated as follows:
\begingroup
\setlength{\abovedisplayskip}{3pt}
\setlength{\belowdisplayskip}{3pt}
\begin{equation} \label{eq:graph-propagation-noisy}
    \tilde{\mathbf{Z}} = \mathbf{\Pi} \cdot \tilde{\mathbf{X}} = \sum\limits_{\ell=0}^{\infty} \alpha (1-\alpha)^{\ell} \cdot (\mathbf{D}^{r-1}\mathbf{A}\mathbf{D}^{-r})^{\ell} \cdot \tilde{\mathbf{X}}.
\end{equation}
\endgroup
We utilize the well-established backward push algorithm~\cite{yin2019scalable} as a standard technique to calculate personalized PageRank. In our work, we extend this algorithm to enable efficient graph propagation.

Algorithm~\ref{alg:graph-propagation} illustrates the pseudo-code of the backward push propagation algorithm. Intuitively, the algorithm starts by setting the residue matrix $\mathbf{R} = \mathbf{D}^{-r}\mathbf{A}$ and the reserve matrix $\mathbf{Q} = \mathbf{0}$ (Line 1). Subsequently, a push procedure (Line 2-6) is executed for each node $v$ if the absolute value of the residue entry $\mathbf{R}(v, j)$ exceeds a threshold $r_{max}$ until no such $v$ exists. Specifically, if there is a node $v$ meeting $|\mathbf{R}(v, j)|>r_{max}$, the algorithm increases the residue of each neighbor $u$ by $(1-\alpha) \cdot \frac{\mathbf{R}(v, j)}{|\mathcal{N}(u)|}$ and increases the reserve of node $v$ by $\alpha \times \mathbf{R}(v, j)$. After that, it reset the residue $\mathbf{R}(v, j)$ to $0$. Finally, the embedding matrix is $\tilde{\mathbf{Z}} = \mathbf{D}^{r} \cdot \mathbf{Q}$. The propagation process can be seen as post-processing and thus does not consume additional privacy budget.

\subsection{Utility Analysis} \label{subsec:utility-analysis}
In this section, we conduct an in-depth theoretical analysis regarding the utility of the PrivGE framework. Additionally, for comparison, we also present a set of theorems that characterize the utility of alternative mechanisms, including the Laplace, Piecewise and Multi-bit mechanisms.

For each node $v \in \mathcal{V}$, we use $\mathbf{z}_{v}$ and $\tilde{\mathbf{z}}_{v}$ to represent the true node embedding vector and the perturbed node embedding vector, respectively. In addition, we define the $\ell_2$ norm of $\mathbf{\Pi}_v$ as $\|\mathbf{\Pi}_v\|_{2} = (\sum_{u\in\mathcal{V}}(\mathbf{\Pi}(v, u))^{2})^{1/2}$ and the $\ell_{\infty}$ norm as $\|\mathbf{\Pi}_v\|_{\infty} = \max_{u\in\mathcal{V}}|\mathbf{\Pi}(v, u)|$. For any dimension $j \in \{1, \dots, d\}$, according to Equations~(\ref{eq:graph-propagation}) and (\ref{eq:graph-propagation-noisy}), we have $z_{v, j} = \sum_{u \in \mathcal{V}} \mathbf{\Pi}(v, u) \cdot x_{u, j}$ and $\tilde{z}_{v, j} = \sum_{u \in \mathcal{V}} \mathbf{\Pi}(v, u) \cdot \tilde{x}_{u, j}$. The following theorem establishes the estimation error of the PrivGE framework.


\begin{theorem} \label{theorem:hds-utility}
    Given $\delta > 0$, for any node $v$, with probability at least $1 - \delta$, we have $ \max_{j \in \{1, \dots, d\}}|\tilde{z}_{v, j}-z_{v, j}| = \mathcal{O}(\log(d / \delta)) $.
\end{theorem}
\begin{proof}
    The variable $\tilde{z}_{v, j} = \sum_{u \in \mathcal{V}} t_{u}$ is the sum of $|\mathcal{V}|$ independent random variables, where $t_{u} = \mathbf{\Pi}(v, u) \cdot \tilde{x}_{u, j}$. According to the Algorithm~{\ref{alg:hds}}, we have $t_{u} \in [a_u, b_u]$, where $a_u = \mathbf{\Pi}(v, u) \cdot (-b-1)$ and $b_u = \mathbf{\Pi}(v, u) \cdot (b+1)$. Observe that $b_u - a_u \leq 2(b+1)$ for any node $u$. Then by Bernstein's inequality, we have
    \begingroup
    \setlength{\abovedisplayskip}{3pt}
    \setlength{\belowdisplayskip}{3pt}
    \begin{align} \label{eq:bernstein}
        & \pr[|\tilde{z}_{v, j} - \sum\nolimits_{u \in \mathcal{V}}\mathbb{E}[t_u]| > \tau] \nonumber \\[-0.2em]
        & < 2 \cdot \exp(-\frac{\tau^{2}}{2\sum_{u \in \mathcal{V}}\var[t_u]+\frac{4}{3}\tau(b+1)}),
    \end{align}
    \endgroup
    where $\mathbb{E}[t_u] = \mathbf{\Pi}(v, u) \cdot \mathbb{E}[\tilde{x}_{u, j}]$ and $\var[t_u] = (\mathbf{\Pi}(v, u))^{2} \cdot \var[\tilde{x}_{u, j}]$.
    The asymptotic expressions involving $\epsilon$ are evaluated in $\epsilon \rightarrow 0$, which yields $ b = \frac{\epsilon / k \cdot e^{\epsilon / k} - e^{\epsilon / k} + 1}{e^{\epsilon / k}(e^{\epsilon / k} - \epsilon / k - 1)} = \mathcal{O}(1) $ and $ \var[\tilde{x}_{u, j}] = \mathcal{O}(k/d) $.
    Therefore, we have
    \begingroup
    \setlength{\abovedisplayskip}{3pt}
    \setlength{\belowdisplayskip}{3pt}
    \begin{align}
        & \pr[|\tilde{z}_{v, j} - \sum\nolimits_{u \in \mathcal{V}}\mathbb{E}[t_u]| > \tau] \nonumber \\[-0.2em]
        & < 2 \cdot \exp(-\frac{\tau^{2}}{\mathcal{O}(k/d) \cdot \sum_{u \in \mathcal{V}} (\mathbf{\Pi}(v, u))^{2} + \tau \cdot \mathcal{O}(1)}).
    \end{align}
    \endgroup
    There exists $\tau = \mathcal{O}(\log(d/\delta))$ such that the following inequality holds with at least $1-\delta/d$ probability
    \begingroup
    \setlength{\abovedisplayskip}{3pt}
    \setlength{\belowdisplayskip}{3pt}
    \begin{align} \label{eq:hds-error-bound-1}
        |\tilde{z}_{v, j} - \sum\nolimits_{u \in \mathcal{V}}\mathbb{E}[t_u]| \leq \tau.
    \end{align}
    \endgroup
    Observe that
    \begingroup
    \setlength{\abovedisplayskip}{3pt}
    \setlength{\belowdisplayskip}{3pt}
    \begin{align} \label{eq:hds-error-bound-2}
        & |\tilde{z}_{v, j} - \sum\nolimits_{u \in \mathcal{V}}\mathbb{E}[t_u]| \nonumber \\[-0.2em]
        &\! = \ |\tilde{z}_{v, j} - z_{v, j} + z_{v, j} - \frac{k}{d}\cdot \frac{b(e^{\epsilon / k} - 1)}{be^{\epsilon / k}+1} \cdot z_{v, j}| \nonumber \\[-0.2em]
        &\! \geq |\tilde{z}_{v, j} - z_{v, j}| - |z_{v, j} - \frac{k}{d}\cdot \frac{b(e^{\epsilon / k} - 1)}{be^{\epsilon / k}+1} \cdot z_{v, j}|.
    \end{align}
    \endgroup
    Since $x_{u, j} \in [-1, 1]$ and $\sum_{u \in \mathcal{V}} \mathbf{\Pi}(v, u) = 1$, combining (\ref{eq:hds-error-bound-2}) in (\ref{eq:hds-error-bound-1}), we have
    \begingroup
    \setlength{\abovedisplayskip}{3pt}
    \setlength{\belowdisplayskip}{3pt}
    \begin{align} \label{eq:hds-error-bound-3}
        |\tilde{z}_{v, j}\! -\! z_{v, j}| \leq \tau\! +\! |z_{v, j}\! -\! \frac{k}{d}\!\cdot\! \frac{b(e^{\epsilon / k} - 1)}{be^{\epsilon / k}+1} \!\cdot\! z_{v, j}| = \mathcal{O}(\log(d/\delta)) \nonumber.
    \end{align}
    \endgroup
    By the union bound, $\max_{j\in\{1, \dots, d\}} |\tilde{z}_{v, j}-z_{v, j}|\! \leq\! \mathcal{O}(\log(d / \delta))$ holds with at least $1 - \delta$ probability.
\end{proof}

Next, we present a series of theorems that delineate the utility of alternative mechanisms.
\begin{theorem} \label{theorem:lp-utility}
    Assume that the perturbation function is the Laplace mechanism. Given $\delta > 0$, for any node $v \in \mathcal{V}$ with probability as least $1 - \delta$, we have
    \begingroup
    \setlength{\abovedisplayskip}{3pt}
    \setlength{\belowdisplayskip}{3pt}
    \begin{align}
        \max_{j \in \{1, \dots, d\}}|\tilde{z}_{v, j}\!-\!z_{v, j}|\! =\! 
        \begin{cases}
        \mathcal{O}(\frac{dlog(d/\delta)}{\epsilon}), \delta < 2de^{-\frac{\|\mathbf{\Pi}_{v}\|_{2}^{2}}{2\|\mathbf{\Pi}_{v}\|_{\infty}^{2}}}, \\
        \mathcal{O}(\frac{d\sqrt{log(d/\delta)}}{\epsilon}), \delta \geq 2de^{-\frac{\|\mathbf{\Pi}_{v}\|_{2}^{2}}{2\|\mathbf{\Pi}_{v}\|_{\infty}^{2}}}.\nonumber
        \end{cases}
    \end{align}
    \endgroup
\end{theorem}

To prove Theorem~\ref{theorem:lp-utility}, our initial step entails demonstrating the sub-exponential nature of the Laplace random variable. Thus, we present the definition of sub-exponential random variables.

\begin{definition}[Sub-exponential distirbutions] \label{def:subE}
    A random variable $\eta$ is said to be sub-exponential with parameter $\nu$ (denoted $\eta \sim subE(\nu)$) if $\mathbb{E}[\eta] = 0$, and its moment generating function (MGF) satisfies
    \begingroup
    \setlength{\abovedisplayskip}{3pt}
    \setlength{\belowdisplayskip}{3pt}
    \begin{align}
        \mathbb{E}[e^{s\eta}] \leq e^{s^2\nu^2/2}, \forall |s|\leq 1 / \nu.
    \end{align}
    \endgroup
\end{definition}

Then, the following lemma confirms that the Laplace distribution is sub-exponential.

\begin{lemma} \label{lemma:lp-subE}
    If a random variable $\eta$ obeys the Laplace distribution with parameter $\lambda$, then $\eta$ is sub-exponential: $\eta \sim subE(2\lambda)$.
\end{lemma}

\begin{proof}
    Without loss of generality, we consider a centered random variable $\eta \sim Lap(1)$. Its moment generating function (MGF) is given by $\mathbb{E}[e^{s\eta}] = \frac{1}{1-s^2}$ for $|s| < 1$. Notably, one of the upper bounds on the MGF is $\mathbb{E}[e^{s\eta}] \leq e^{2s^2}$ for $|s| < \frac{1}{2}$. This indicates that $\eta \sim subE(2)$.

    Next, we extend the result to the Laplace distribution with parameter $\lambda$. Note that if $\eta \sim Lap(1)$, then $\lambda \eta \sim Lap(\lambda)$. As a result, the upper bound on the MGF becomes $\mathbb{E}[e^{s\lambda\eta}] \leq e^{2\lambda^2s^2}$ for $|s|<\frac{1}{2\lambda}$, from which we conclude that the distribution $Lap(\lambda)$ is sub-exponential with parameter $\nu=2\lambda$.
\end{proof}
Now, we prove Theorem~\ref{theorem:lp-utility}.

\begin{proof}
    If the perturbation function is the Laplace mechanism, we have $\tilde{z}_{v, j}-z_{v, j} = \sum_{u\in \mathcal{V}}\mathbf{\Pi}(v, u)\cdot \eta_{u}$, where $\eta_{u} \sim Lap(2d/\epsilon)$. Since the random variable $\eta_u \sim subE(4d/\epsilon)$, according to the Bernstein's inequality we have
    \begingroup
    \setlength{\abovedisplayskip}{3pt}
    \setlength{\belowdisplayskip}{3pt}
    \begin{align}
        & \pr[|\tilde{z}_{v, j}-z_{v, j}| > \tau] \nonumber = \pr[|\sum\nolimits_{u\in \mathcal{V}}\mathbf{\Pi}(v, u)\cdot \eta_{u}| > \tau] \nonumber \\[-0.2em]
        & < \begin{cases}
            2e^{-\tau^2\epsilon^2/32d^2\|\mathbf{\Pi}_v\|_{2}^{2}}, & \text{if} \ 0 \leq \tau \leq \frac{4d\|\mathbf{\Pi}_v\|_{2}^{2}}{\epsilon\|\mathbf{\Pi}_v\|_{\infty}}, \\[-0.2em]
            2e^{-\tau\epsilon/8d\|\mathbf{\Pi}_v\|_{\infty}}, & \text{if} \ \tau > \frac{4d\|\mathbf{\Pi}_v\|_{2}^{2}}{\epsilon\|\mathbf{\Pi}_v\|_{\infty}}.
        \end{cases}
    \end{align}
    \endgroup

    First, consider the case where $0 \leq \tau \leq \frac{4d\|\mathbf{\Pi}_v\|_{2}^{2}}{\epsilon\|\mathbf{\Pi}_v\|_{\infty}}$. Let $\delta/d=2e^{-\tau^2\epsilon^2/32d^2\|\mathbf{\Pi}_v\|_{2}^{2}}$. Solving for $\tau$, we obtain $ \tau=\frac{4d}{\epsilon}\cdot \sqrt{2log(2d/\delta)}\cdot \|\mathbf{\Pi}_{v}\|_{2} $.
    In this case, $\delta$ must satisfy the condition $\delta \geq 2de^{-\frac{\|\mathbf{\Pi}_{v}\|_{2}^{2}}{2\|\mathbf{\Pi}_{v}\|_{\infty}^{2}}}$. Utilizing the union bound and evaluating the asymptotic expressions for small $\epsilon$ (i.e., $\epsilon \rightarrow 0$), we can deduce that when $\delta \geq 2de^{-\frac{\|\mathbf{\Pi}_{v}\|_{2}^{2}}{2\|\mathbf{\Pi}_{v}\|_{\infty}^{2}}}$, there exists $\tau=\mathcal{O}(d\sqrt{log(d/\delta)}/\epsilon)$ such that the inequality $\max_{j \in \{1, \dots, d\}}|\tilde{z}_{v, j}-z_{v, j}|\leq\tau$ holds with at least $1 - \delta$ probability.

    Second, suppose $\tau \geq \frac{4d\|\mathbf{\Pi}_v\|_{2}^{2}}{\epsilon\|\mathbf{\Pi}_v\|_{\infty}}$. Similar to the first case, let $\delta/d=2e^{-\tau\epsilon/8d\|\mathbf{\Pi}_v\|_{\infty}}$. Solving the above for $\tau$, we have $ \tau=\frac{8d}{\epsilon}\cdot log(2d/\delta)\cdot \|\mathbf{\Pi}_{v}\|_{\infty} $.
    For this case to be valid, $\delta$ must satisfy $\delta < 2de^{-\frac{\|\mathbf{\Pi}_{v}\|_{2}^{2}}{2\|\mathbf{\Pi}_{v}\|_{\infty}^{2}}}$. Similarly, according to union bound, when $\delta < 2de^{-\frac{\|\mathbf{\Pi}_{v}\|_{2}^{2}}{2\|\mathbf{\Pi}_{v}\|_{\infty}^{2}}}$, there exists $\tau=\mathcal{O}(dlog(d/\delta)/\epsilon)$ such that the inequality $\max_{j \in \{1, \dots, d\}}|\tilde{z}_{v, j}-z_{v, j}|\leq\tau$ holds with at least $1 - \delta$ probability.
\end{proof}

\begin{theorem} \label{theorem:pm-mb-utility}
    Assume that the perturbation function is the Piecewise mechanism or the Multi-bit mechanism. Given $\delta > 0$, for any node $v$, with probability at least $1 - \delta$, we have $ \max_{j \in \{1, \dots, d\}}|\tilde{z}_{v, j}-z_{v, j}| = \mathcal{O}(\frac{d \log(d/\delta)}{\epsilon}) $.
\end{theorem}

\begin{proof}
    If the perturbation function is Piecewise mechanism, for any node $u$, $\mathbf{\Pi}(v, u) \cdot (\tilde{x}_{u, j} - x_{u, j})$ is a zero-mean random variable and its variance is $(\mathbf{\Pi}(v, u))^2\var[\tilde{x}_{u, j}]$. Besides, the inequality $|\mathbf{\Pi}(v, u) \cdot (\tilde{x}_{u, j} - x_{u, j})| \leq \frac{2de^{\epsilon/2k}}{k(e^{\epsilon/2k}-1)}$ always holds. Then by Bernstein's inequality, we find that
    \begingroup
    \setlength{\abovedisplayskip}{3pt}
    \setlength{\belowdisplayskip}{3pt}
    \begin{align} \label{eq:pm-bernstein}
        & \pr[|\tilde{z}_{v, j}-z_{v, j}| > \tau] = \pr[|\sum\nolimits_{u \in \mathcal{V}}\mathbf{\Pi}(v, u) \cdot (\tilde{x}_{u, j} - x_{u, j})| > \tau] \nonumber \\[-0.2em]
        & < 2 \exp(\frac{-\tau^{2}}{2\sum_{u \in \mathcal{V}}(\mathbf{\Pi}(v, u))^{2}\var[\tilde{x}_{u, j}]+\frac{\tau \cdot 4de^{\epsilon/2k}}{3k(e^{\epsilon/2k}-1)}}).
    \end{align}
    \endgroup
    Note that asymptotic expressions involving $\epsilon$ are in the sense of $\epsilon \gets 0$. Thus, we can deduce that $ \var[\tilde{x}_{u, j}]=\mathcal{O}(\frac{kd}{\epsilon^{2}}) $ and $ \frac{de^{\epsilon/2k}}{k(e^{\epsilon/2k}-1)} = \mathcal{O}(\frac{d}{\epsilon}) $.
    Since $\mathbf{\Pi}(v, u) \in [0, 1]$ and $\sum_{u \in \mathcal{V}} \mathbf{\Pi}(v, u) = 1$, we can obtain that
    \begingroup
    \setlength{\abovedisplayskip}{3pt}
    \setlength{\belowdisplayskip}{3pt}
    \begin{align}
        \pr[|\tilde{z}_{v, j}-z_{v, j}| > \tau] < 2 \cdot \exp(\frac{-\tau^{2}}{\mathcal{O}(\frac{kd}{\epsilon^{2}})+\tau \cdot \mathcal{O}(\frac{d}{\epsilon})}).
    \end{align}
    \endgroup
    By applying the union bound, we can ensure that $\max_{j \in \{1, \dots, d\}}|\tilde{z}_{v, j}-z_{v, j}| \leq \tau$ holds with at least $1-\delta$ probability by setting $ \delta / d = 2 \cdot \exp(\frac{-\tau^{2}}{\mathcal{O}(\frac{kd}{\epsilon^{2}})+\tau \cdot \mathcal{O}(\frac{d}{\epsilon})}) $.
    Solving the above for $\tau$, we have $\tau = \mathcal{O}(\frac{d \log(d/\delta)}{\epsilon})$.

    Given that the Multi-bit mechanism belongs to the same category as the Piecewise mechanism (both being bounded mechanisms) and the perturbed value is also unbiased, the analysis process of the Multi-bit mechanism closely resembles that of the Piecewise mechanism. Due to space constraints, we omit the proof of the Multi-bit mechanism.
\end{proof}

According to the utility analysis presented in Theorem~\ref{theorem:hds-utility} and Theorem~\ref{theorem:lp-utility}, we observe that the HDS mechanism can yield a higher utility than the Laplace mechanism. Furthermore, as indicated by Theorem~\ref{theorem:hds-utility} and Theorem~\ref{theorem:pm-mb-utility}, the proposed HDS mechanism offers superior error bounds compared to the Piecewise and Multi-bit mechanisms, reducing them from $\mathcal{O}(\frac{d\log(d/\delta)}{\epsilon})$ to $\mathcal{O}(\log(d/\delta)$).

\section{Experiments} \label{sec:experiment}
\subsection{Experimental Setup} \label{subsec:exp-setup}
\paragraph{Dataset}
Our experiments use five real-world datasets: two citation networks (i.e., Cora and Pubmed~\cite{kipf2016semi}), one social networks (i.e., LastFM~\cite{sajadmanesh2021locally}) and two web graphs (i.e., Facebook and Wikipedia~\cite{rozemberczki2021multi}). Table~\ref{tab:datasets} summarizes the statistics of the datasets used for evaluation. Following previous works~\cite{sajadmanesh2021locally,lin2022towards}, we randomly split each dataset into three portions for node classification: $ 50\% $ for training, $ 25\% $ for validation, and $ 25\% $ for testing. Since the labels of nodes in the LastFM dataset are highly imbalanced, we restrict the classes to the top $ 10 $ with the most samples. For link prediction, the edges are divided into $ 85\% $ for training, $ 10\% $ for testing, and $ 5\% $ for validation. And we sample the same number of non-existing links in each group as negative links. Note that the LDP mechanisms perturb the node features of all training, validation, and test sets.
\paragraph{Model Implementation Details}
For node classification, we first obtain the node embedding matrix $\tilde{\mathbf{Z}}$ via the PrivGE framework. Then the matrix $\tilde{\mathbf{Z}}$ is fed into a model composed of a multi-layer perceptron (MLP) followed by a softmax function for prediction:
\begingroup
\setlength{\abovedisplayskip}{3pt}
\setlength{\belowdisplayskip}{3pt}
\begin{align}
    \hat{\mathbf{Y}} = \text{softmax}(\text{MLP}(\tilde{\mathbf{Z}}; \mathbf{\Theta})),
\end{align}
\endgroup
where $\hat{\mathbf{Y}}$ represents the posterior class probabilities and $\mathbf{\Theta}$ is the learnable model parameters. Moreover, we adopt the cross-entropy loss as the optimization function to train the classification module.

\begin{table}[!t]
    \footnotesize
    \centering
    \caption{Dataset Statistics}
    \label{tab:datasets}
    \begin{tabular}{ccccc}
        \toprule
        \textsc{Dataset} & \textsc{\#Nodes} & \textsc{\#Edges} & \textsc{\#Features} & \textsc{\#Classes} \\ 
        \midrule
        \textsc{Cora} & 2,708 & 5,278 & 1,433 & 7 \\ 
        \textsc{Pubmed} & 19,717 & 44,324 & 500 & 3 \\ 
        \textsc{Lastfm} & 7,083 & 25,814 & 7,842 & 10 \\ 
        \textsc{Facebook} & 22,470 & 170,912 & 4,714 & 4 \\ 
        \textsc{Wikipedia} & 11,631 & 170,845 & 13,183 & 2 \\ 
        \bottomrule
    \end{tabular}
    \vspace*{-0.1cm}
\end{table}

To perform link prediction, we first extract edge features based on node representations. Specifically, for each node pair $(u, v) \in \mathcal{E}$, we combine the node vectors $\tilde{\mathbf{z}}_{u}$ and $\tilde{\mathbf{z}}_{v}$ via the Hadamard product to compose the edge vector. Then, we take the constructed edge feature vectors as input and train a logistic regression classifier to predict the presence or absence of edges. Following existing works~\cite{wang2021pairwise,lv2022path}, we use the pairwise objective loss function to optimize the model.

For each task, we train the model on the training set and tune the hyperparameters based on the model's performance in the validation set. First, we define the search space of hyperparameters. Then we fix the privacy budget $\epsilon = 1.0$ and utilize NNI~\cite{nni2021}, an automatic hyperparameter optimization tool, to determine the optimal choices. We use the Adam optimizer to train the models and select the best model to evaluate the testing set.

\paragraph{Competitors and Evaluation Metric}
In our experiments, we compare the performance of the HDS mechanism with the Laplace mechanism (LP)~\cite{dwork2006calibrating}, the Piecewise mechanism (PM)~\cite{wang2019collecting} and the Multi-bit mechanism (MB)~\cite{sajadmanesh2021locally}. Following existing works~\cite{kipf2016semi,sajadmanesh2021locally}, we use \textit{accuracy} to evaluate the node classification performance and \textit{AUC} to evaluate the link prediction performance. Unless otherwise stated, we run each algorithm ten times for both tasks and report the mean and standard deviation values.

\paragraph{Software and Hardware}
We implement our algorithm in PyTorch and C++. All the experiments are conducted on a machine with an NVIDIA GeForce RTX 3080Ti, an AMD Ryzen Threadripper PRO 3995WX CPU, and 256GB RAM. Our code is available online\footnote[3]{\url{https://github.com/Zening-Li/PrivGE}}.

\subsection{Experimental Results} \label{subsec:exp-res}
\begin{figure*}[!t]
    \centering
    \subfloat[Cora]{\includegraphics[width=0.42\columnwidth, height=2.3cm]{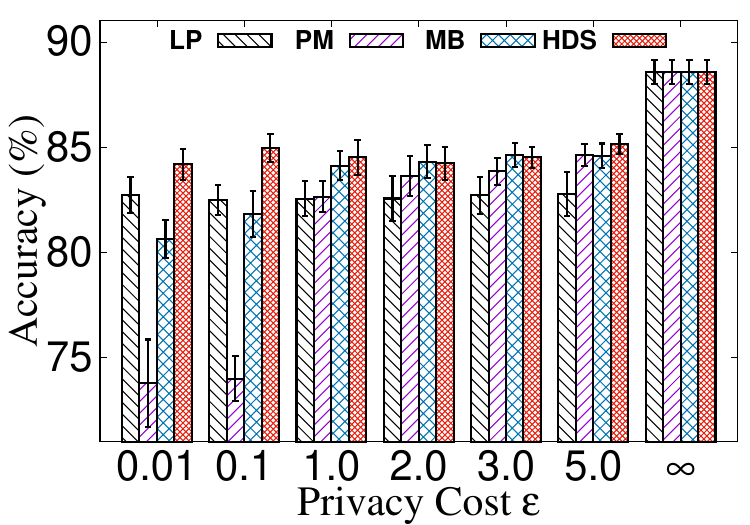}}
    \subfloat[Pubmed]{\includegraphics[width=0.42\columnwidth, height=2.3cm]{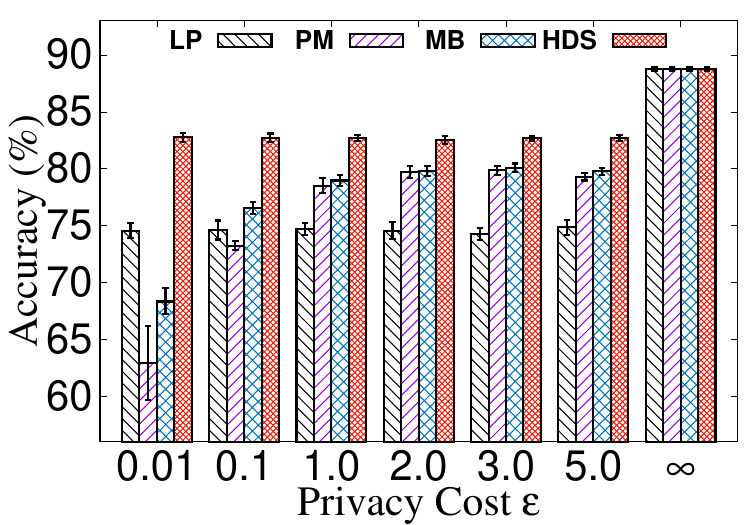}}
    \subfloat[LastFM]{\includegraphics[width=0.42\columnwidth, height=2.3cm]{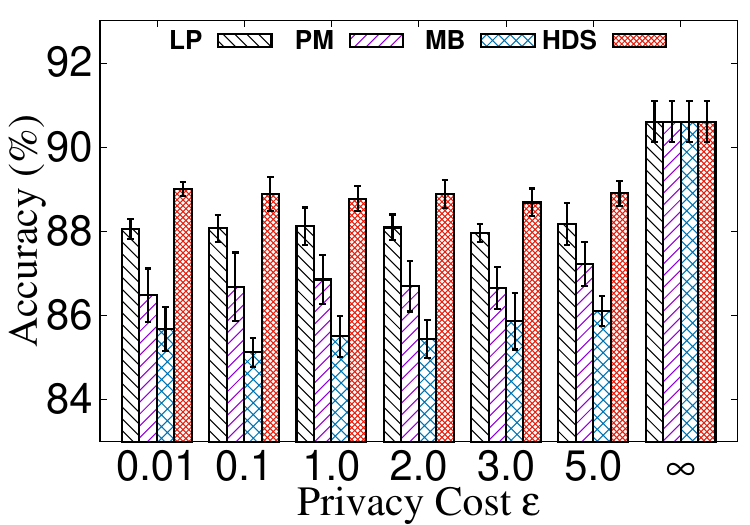}}
    \subfloat[Facebook]{\includegraphics[width=0.42\columnwidth, height=2.3cm]{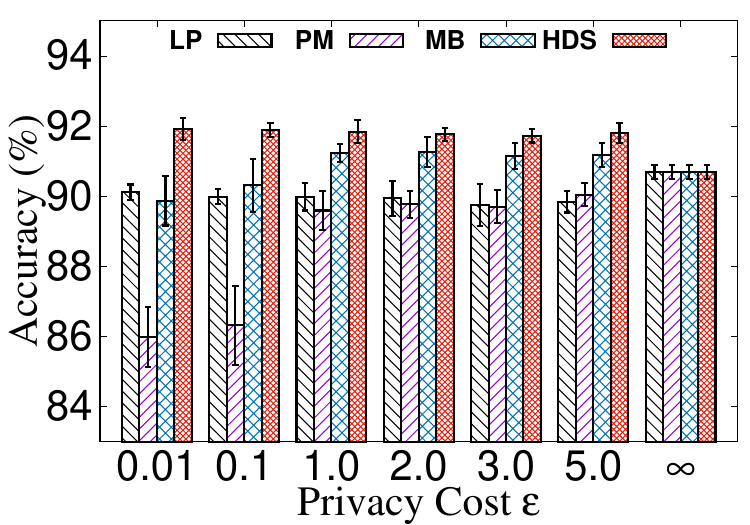}}
    \subfloat[Wikipedia]{\includegraphics[width=0.42\columnwidth, height=2.3cm]{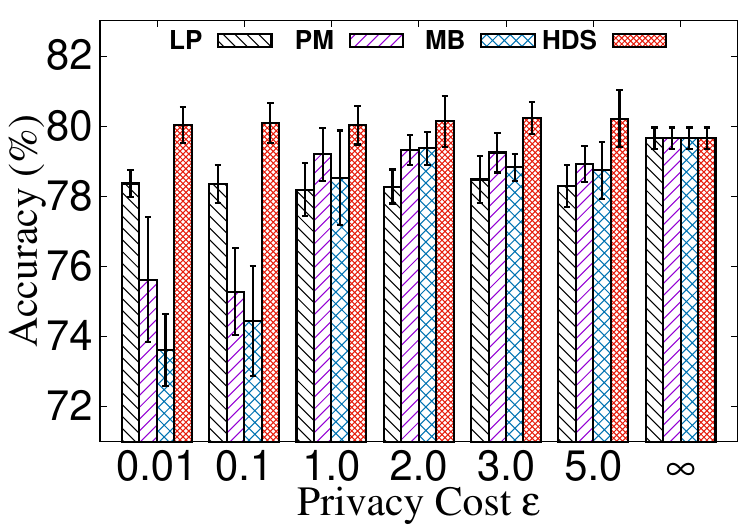}}
    \caption{Trade-offs between privacy and accuracy under different LDP mechanisms in node classification. Note that the error bars represent the standard deviation and the results for $\infty$ denote the accuracy of the non-private baselines.}
    \label{fig:node-classification}
    \vspace*{-0.1cm}
\end{figure*}
\paragraph{Node Classification}
In the first set of experiments, we evaluate the performance of different mechanisms under different privacy budgets for node classification. The privacy budget varies over the range $\{0.01, 0.1, 1.0, 2.0, 3.0, 5.0, \infty\}$, and we report the accuracy of each method under each privacy budget, as shown in Figure~\ref{fig:node-classification}. Note that the error bars in the Figure~\ref{fig:node-classification} represent the standard deviation. In addition, the case where $\epsilon = \infty$ is provided for comparison with non-private baselines, where node features are directly employed without any perturbation.

The experimental results lead us to conclude that our proposed method demonstrates robustness to the perturbations and achieves comparable performance to the non-private baseline across all datasets. For instance, on the LastFM dataset, the HDS mechanism achieves an accuracy of about $ 89.0\% $ at $\epsilon=0.01$, with only a $ 1.6\% $ decrease compared to the non-private method ($\epsilon=\infty$). On the Cora dataset, the accuracy of HDS mechanism at $ \epsilon=0.01 $ is approximately $ 84.2\% $, just $ 4.3\% $ lower than the non-private method. Similarly, on the Pubmed dataset, when $ \epsilon=0.01 $, our method loses less than $ 6.0\% $ in accuracy over the non-private baseline. Interestingly, on the Facebook and Wikipedia datasets, our framework outperforms the non-private setting by about $ 1.2 $ and $ 0.4 $ percentage points, respectively. We conjecture that the observed experimental results are mainly due to the fact that the injected noise provides the model with strong generalization capabilities. Additionally, the small standard deviation in our experimental results indicates that our method maintains stable performance under various privacy budgets. It consistently exhibits similar accuracy across distinct privacy budget values from $ 0.01 $ to $ 1.0 $, which serves as a powerful demonstration of its robustness to perturbations.

Second, our HDS mechanism consistently outperforms the other mechanisms in almost all cases, particularly under smaller privacy budgets, which is consistent with the theoretical analysis in Theorem~\ref{theorem:hds-utility}. For instance, at $ \epsilon=0.01 $, HDS achieves approximately $ 8.2\% $ higher accuracy than the best competitor, the Laplace mechanism, on the Pubmed dataset. Similarly, at $ \epsilon=0.01 $, our approach outperforms LP, PM, and MB by $ 1.7\% $, $ 4.4\% $ and $ 6.4\% $, respectively, on the Wikipedia dataset. This remarkable performance advantage can be attributed to the fact that our proposed perturbation mechanism can provide more concentrated perturbations compared to the Laplace, Piecewise, and Multi-bit mechanisms. Additionally, we can observe that even the simplest method, the Laplace mechanism, sometimes outperforms the Piecewise and Multi-bit mechanisms, especially when a smaller privacy budget is allocated.

\begin{table}[!t]
    \footnotesize
    \centering
    \caption{Trade-offs between privacy and AUC under different LDP mechanisms in link prediction. Note that the results for $\infty$ denote the AUC of the non-private baselines.}
    \label{tab:link-prediction}
    \begin{tabular}{cccccc}
        \toprule
        \textsc{Dataset} & \textsc{Mech.} & $\epsilon=1.0$ & $\epsilon=2.0$ & $\epsilon=3.0$ & $\epsilon=5.0$ \\ 
        \midrule
        \multirow{2}*{\textsc{Cora}}
        & LP & $75.2 \pm 3.1$ & $76.0 \pm 3.5$ & $75.0 \pm 3.7$ & $75.2 \pm 3.6$ \\ 
        & PM & $73.2 \pm 3.2$ & $78.2 \pm 1.1$ & $77.8 \pm 1.7$ & $75.2 \pm 2.1$ \\ 
        \multirow{1}*{$ \epsilon=\infty $}
        & MB & $76.0 \pm 1.8$ & $78.5 \pm 1.4$ & $78.5 \pm 1.1$ & $76.6 \pm 0.9$ \\ 
        \multirow{1}*{$93.1 \pm 0.3$}
        & HDS & $\bm{82.4} \pm \bm{1.5}$ & $\bm{82.7} \pm \bm{0.8}$ & $\bm{82.5} \pm \bm{0.9}$ & $\bm{82.3} \pm \bm{0.8}$ \\ 
        \midrule
        \multirow{2}*{\textsc{Pubmed}}
        & LP & $62.8 \pm 0.6$ & $63.2 \pm 1.2$ & $64.3 \pm 0.8$ & $64.4 \pm 0.5$ \\ 
        & PM & $79.0 \pm 1.0$ & $77.0 \pm 1.2$ & $76.7 \pm 1.5$ & $74.9 \pm 0.5$ \\ 
        \multirow{1}*{$ \epsilon=\infty $}
        & MB & $79.4 \pm 0.8$ & $77.9 \pm 1.0$ & $78.4 \pm 0.7$ & $77.2 \pm 0.7$ \\ 
        \multirow{1}*{$97.5 \pm 0.1$}
        & HDS & $\bm{79.5} \pm \bm{1.5}$ & $\bm{79.6} \pm \bm{1.6}$ & $\bm{80.1} \pm \bm{0.2}$ & $\bm{79.6} \pm \bm{1.5}$ \\ 
        \midrule
        \multirow{2}*{\textsc{Lastfm}}
        & LP & $74.4 \pm 6.3$ & $75.4 \pm 6.0$ & $77.0 \pm 4.8$ & $73.8 \pm 3.5$ \\ 
        & PM & $67.9 \pm 0.9$ & $71.7 \pm 6.2$ & $72.0 \pm 3.9$ & $80.1 \pm 1.4$ \\ 
        \multirow{1}*{$ \epsilon=\infty $}
        & MB & $75.7 \pm 3.1$ & $78.2 \pm 3.3$ & $78.9 \pm 4.6$ & $78.0 \pm 3.3$ \\ 
        \multirow{1}*{$95.9 \pm 0.1$}
        & HDS & $\bm{91.7} \pm \bm{1.3}$ & $\bm{92.0} \pm \bm{0.1}$ & $\bm{92.0} \pm \bm{0.1}$ & $\bm{92.0} \pm \bm{0.1}$ \\ 
        \midrule
        \multirow{2}*{\textsc{Facebook}}
        & LP & $81.92 \pm 3.2$ & $84.7 \pm 2.7$ & $85.6 \pm 1.1$ & $86.0 \pm 1.2$ \\ 
        & PM & $92.1 \pm 0.5$ & $93.2 \pm 0.5$ & $93.3 \pm 0.3$ & $89.9 \pm 0.3$ \\ 
        \multirow{1}*{$ \epsilon=\infty $}
        & MB & $92.6 \pm 3.8$ & $92.8 \pm 2.3$ & $93.1 \pm 2.2$ & $87.7 \pm 0.9$ \\ 
        \multirow{1}*{$95.6 \pm 0.1$}
        & HDS & $\bm{96.7} \pm \bm{0.1}$ & $\bm{96.7} \pm \bm{0.1}$ & $\bm{96.6} \pm \bm{0.1}$ & $\bm{96.6} \pm \bm{0.1}$ \\ 
        \midrule
        \multirow{2}*{\textsc{Wikipedia}}
        & LP & $72.2 \pm 2.8$ & $74.9 \pm 4.7$ & $76.1 \pm 2.5$ & $75.7 \pm 3.5$ \\ 
        & PM & $76.9 \pm 5.3$ & $76.6 \pm 4.6$ & $76.8 \pm 3.5$ & $77.9 \pm 1.7$ \\ 
        \multirow{1}*{$ \epsilon=\infty $}
        & MB & $76.6 \pm 3.6$ & $77.2 \pm 1.4$ & $78.2 \pm 1.3$ & $78.4 \pm 1.1$ \\ 
        \multirow{1}*{$98.5 \pm 0.1$}
        & HDS & $\bm{99.1} \pm \bm{0.1}$ & $\bm{99.1} \pm \bm{0.1}$ & $\bm{99.1} \pm \bm{0.1}$ & $\bm{99.1} \pm \bm{0.1}$ \\ 
        \bottomrule
    \end{tabular}
\end{table}

\paragraph{Link Prediction}
In the second set of experiments, we shift our focus to another important task: link prediction. To examine the performance of PrivGE under different privacy budgets, we vary $\epsilon$ from $1.0$ to $5.0$. The experimental results are summarized in Table~\ref{tab:link-prediction}.

\begin{figure*}[!t]
	\centering
    \subfloat[Cora]{\includegraphics[width=0.37\columnwidth, height=2.2cm]{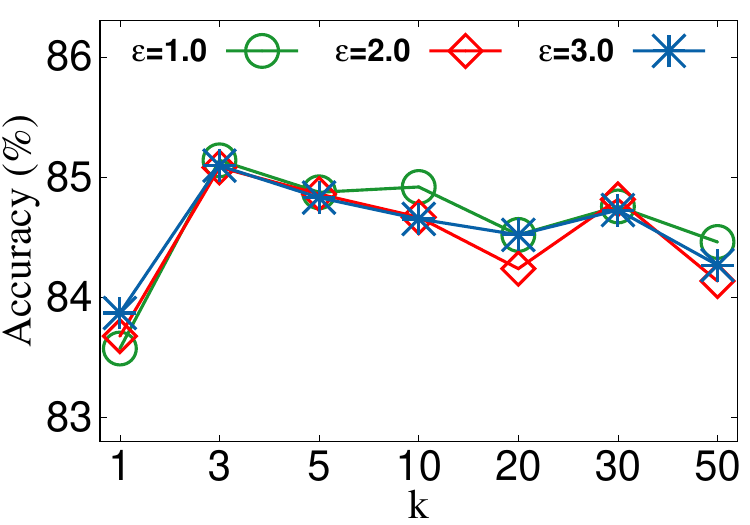}} \hspace*{0.15cm}
    \subfloat[Pubmed]{\includegraphics[width=0.37\columnwidth, height=2.2cm]{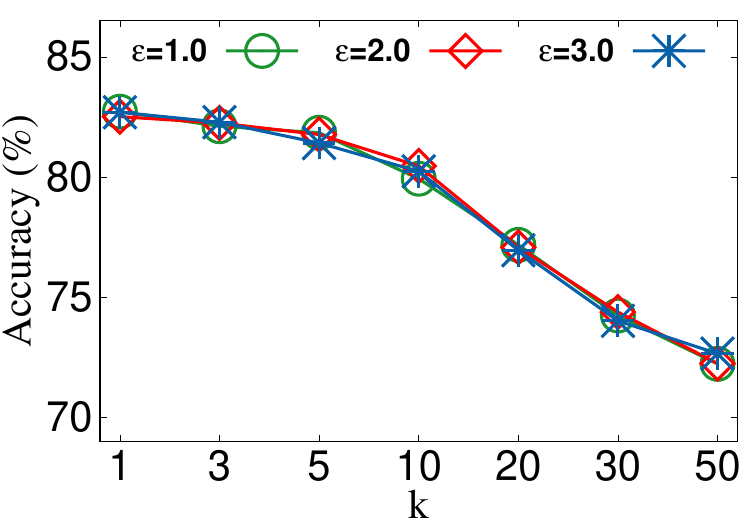}} \hspace*{0.15cm}
    \subfloat[LastFM]{\includegraphics[width=0.37\columnwidth, height=2.2cm]{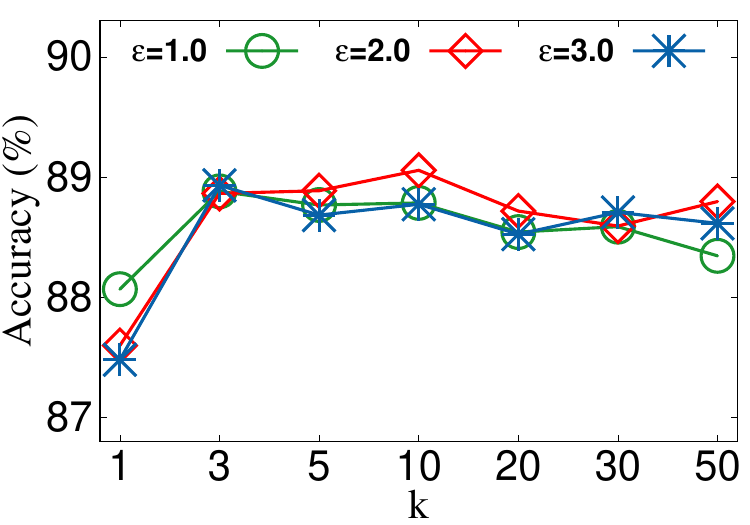}} \hspace*{0.15cm}
    \subfloat[Facebook]{\includegraphics[width=0.37\columnwidth, height=2.2cm]{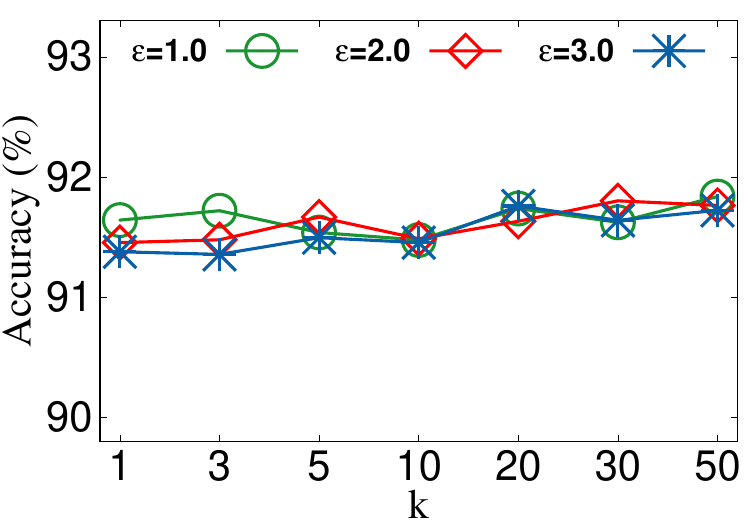}} \hspace*{0.15cm}
    \subfloat[Wikipedia]{\includegraphics[width=0.37\columnwidth, height=2.2cm]{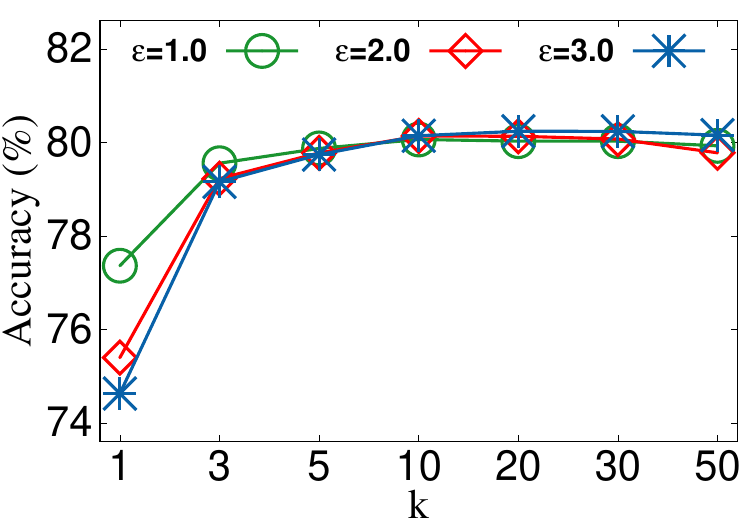}}
	\caption{Effect of the sampling parameter $k$ on the performance of PrivGE for node classification.}
	\label{fig:sampling-parameter-node}
    \Description{effect of k on node classification}
    \vspace*{-0.3cm}
\end{figure*}

\begin{figure*}[t!]
    \vspace*{-0.25cm}
	\centering
    \subfloat[Cora]{\includegraphics[width=0.37\columnwidth, height=2.2cm]{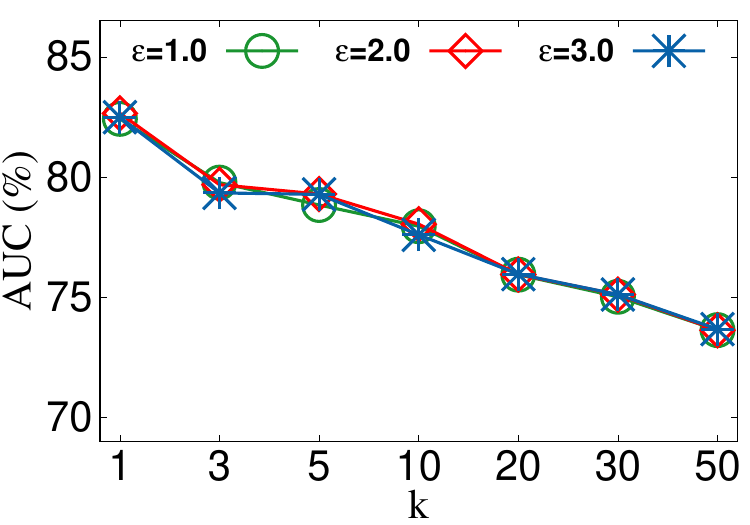}} \hspace*{0.15cm}
    \subfloat[Pubmed]{\includegraphics[width=0.37\columnwidth, height=2.2cm]{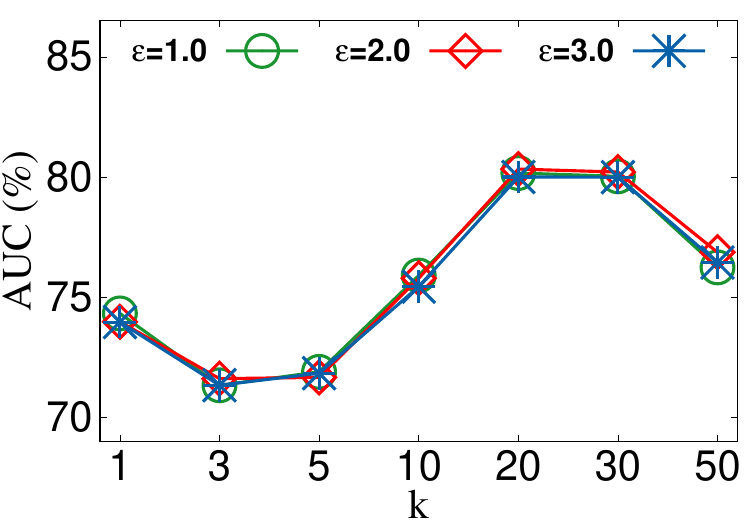}} \hspace*{0.15cm}
    \subfloat[LastFM]{\includegraphics[width=0.37\columnwidth, height=2.2cm]{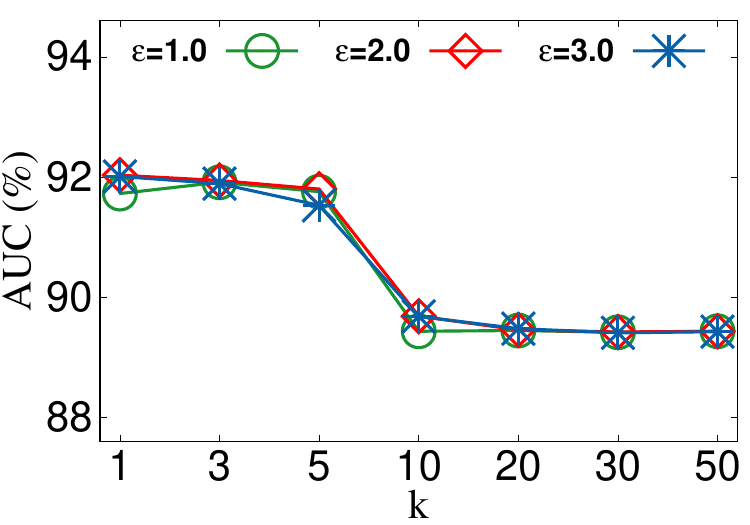}} \hspace*{0.15cm}
    \subfloat[Facebook]{\includegraphics[width=0.37\columnwidth, height=2.2cm]{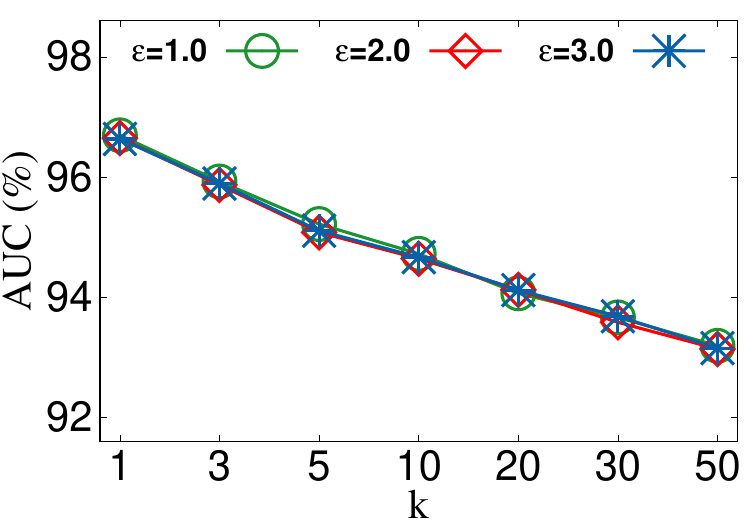}} \hspace*{0.15cm}
    \subfloat[Wikipedia]{\includegraphics[width=0.37\columnwidth, height=2.2cm]{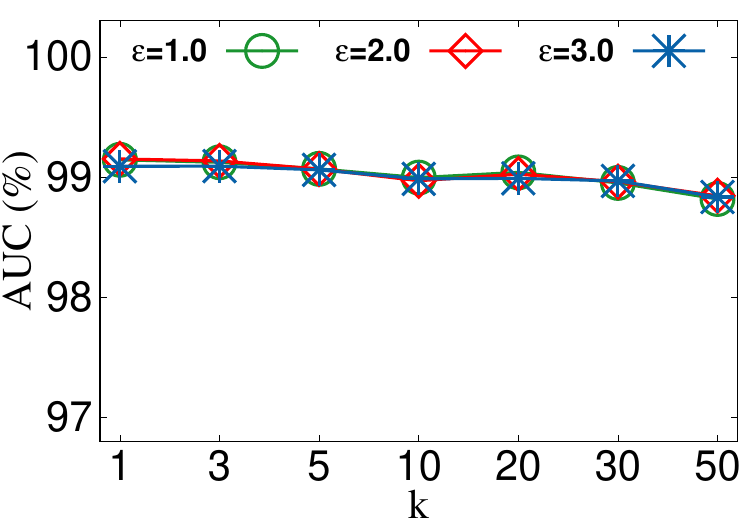}}
	\caption{Effect of the sampling parameter $k$ on the performance of PrivGE for link prediction.}
	\label{fig:sampling-parameter-link}
    \Description{effect of k on link prediction}
    \vspace*{-0.2cm}
\end{figure*}

Similar to the node classification task, the HDS mechanism outperforms the other three methods in terms of the AUC score. At $\epsilon=1.0$, the AUC scores of our proposed mechanism outperform the best competitors by about $6.5$, $16.0$, and $4.0$ on the datasets Cora, LastFM, and Facebook, respectively. In addition, even considering strong privacy guarantees, such as $\epsilon=1.0$, the technique still achieves acceptable AUC scores on all five datasets, especially on the LastFM, Facebook and Wikipedia datasets, where the AUC scores are above $90\%$. These outcomes underline the effectiveness of the proposed method for the link prediction task. Similar to the node classification task, our perturbation mechanism yields higher AUC scores than the non-private baseline on both Facebook and Wikipedia datasets, which further confirms the efficiency of the proposed solution. Unlike the observations obtained from the node classification, in the link prediction task, we notice that the Laplace mechanism is inferior to the Piecewise and Multi-bit mechanisms in most cases. In addition, the performance of the three competitors becomes very unstable since a lot of noise is injected. As a result, in some cases, the performance with a small privacy budget yet outperforms when the privacy budget is large. In summary, the experimental results emphasize that our approach can achieve a better trade-off between privacy and utility.

\paragraph{Parameter Analysis}
In this experiment, we investigate the impact of the parameter $k$ on the performance of our proposed method. The parameter $ k $ is a hyperparameter that controls the size of the subset of dimensions that are randomly perturbed in the HDS mechanism. We vary $ k $ within $\{1, 3, 5, 10, 20, 30, 50\}$, and evaluate the results under different privacy budgets $\epsilon \in \{1.0, 2.0, 3.0\}$ for both node classification and link prediction tasks.

As for node classification, the experimental results are illustrated in Figure~\ref{fig:sampling-parameter-node}. We observe that the performance of our mechanism remains relatively stable for all values of $k$ on Cora, LastFM, and Facebook datasets, varying up and down by no more than $ 2.0\% $. However, on the Pubmed dataset, the approach performs better when $ k $ is smaller, especially when $k \leq 5$. In contrast, on the Wikipedia dataset, the performance of HDS improves as the number of samples increases, particularly when $k \geq 10 $. This indicates that the performance of HDS is sensitive to the value of $k$ on the Pubmed and Wikipedia datasets, whereas it is relatively insensitive to $k$ on the other three datasets.

As for link prediction, the experimental results are illustrated in Figure~\ref{fig:sampling-parameter-link}. Our framework performs better for small values of $k$ on Cora, LastFM, and Facebook datasets, while it achieves better results for large $k$ values on the Pubmed dataset. On the Wikipedia dataset, our method is stable across all values of $k$. Generally, the optimal value of $k$ depends on the specific dataset and task.

\section{Related work} \label{sec:related-work}
\paragraph{DP on Graph Analysis}
DP has become the standard for privacy protection in various data analysis tasks~\cite{kasiviswanathan2011can, duchi2013local}. In centralized DP, most graph analysis tasks revolve around the computation of diverse statistics such as degree distributions and subgraph counts~\cite{nissim2007smooth, karwa2011private, kasiviswanathan2013analyzing, zhang2015private, day2016publishing, ding2021differentially}. In addition, research extends to other graph problems under DP, such as node subset release for vertex cover~\cite{gupta2010differentially} and densest subgraph~\cite{nguyen2021differentially, dhulipala2022differential}, and synthetic graph generation~\cite{jorgensen2016publishing}. However, all these centralized DP methods require possession of all user data, which suffers from the data breach.

Local DP assumes an untrusted data collector and recently has attracted much attention in graph analysis. Ye et al.~\cite{ye2020towards} propose a method to estimate graph metrics. Qin et al.~\cite{qin2017generating} develop a multi-phase approach to generate synthetic decentralized social networks under the notion of LDP. In addition, a series of studies have focused on subgraph counting~\cite{sun2019analyzing, imola2021locally, imola2021communication}. Sun et al.~\cite{sun2019analyzing} develop a multi-phase framework under decentralized DP, which assumes that each user allows her friends to see all her connections. Imola et al.~\cite{imola2021locally} introduce an additional round of interaction between users and the data collector to reduce the estimation error, and~\cite{imola2021communication} employs edge sampling to improve the communication efficiency. 

\paragraph{DP on Graph Learning}
To address the privacy concerns in graph learning, DP has been widely used to protect sensitive data. For instance, Xu et al.~\cite{xu2018dpne} propose a DP algorithm for graph embedding that uses the objective perturbation mechanism on the loss function of the matrix factorization. Zhang et al.~\cite{zhang2019graph} devise a perturbed gradient descent method based on the Lipschitz condition. Epasto et al.~\cite{epasto2022differentially} develop an approximate method for computing personalized PageRank vectors with differential privacy and extend this algorithm to graph embedding. In recent years, DP has also been used to provide formal privacy assurances for Graph Neural Networks~\cite{daigavane2021node, kolluri2022lpgnet, sajadmanesh2023gap}. All these methods assume that there is a trusted data curator, making them susceptible to data leakage issues and unsuitable for decentralized graph analysis applications.

To tackle these issues, some studies have focused on leveraging LDP to train Graph Neural Networks~\cite{sajadmanesh2021locally, zhang2021graph, lin2022towards, jin2022gromov}. To be specific, Zhang et al.~\cite{zhang2021graph} propose an algorithm for recommendation systems, which utilizes LDP to prevent users from attribute inference attacks. LPGNN~\cite{sajadmanesh2021locally} assumes that node features are private and the data curator has access to the graph structure, where the scenario is similar to ours. In this setting, each user perturbs their features through the Multi-bit mechanism. However, this mechanism introduces much noise to the data that can degrade the performance. Analogously,~\cite{lin2022towards,jin2022gromov} also utilize the Multi-bit mechanism to preserve the privacy of node features. Unlike these works, we propose an improved mechanism to achieve LDP for graph embedding and provide a detailed utility analysis for our method.
\section{Conclusion} \label{sec:conclusion}
In this paper, we propose PrivGE, a privacy-preserving graph embedding framework based on local differential privacy. To this end, we propose an LDP mechanism called HDS to protect the privacy of node features. Then, to avoid neighborhood explosion and over-smoothing problems, we decouple the feature transformation from graph propagation and employ personalized PageRank as a proximity measure to learn node representations. Importantly, we perform a novel and comprehensive theoretical analysis of the privacy and utility of the PrivGE framework. Extensive experiments conducted on real-world datasets demonstrate that our proposed method establishes state-of-the-art performance and achieves better privacy-utility trade-offs on both node classification and link prediction tasks. In future work, we plan to extend our work to protect the privacy of the graph structure as well. Another future direction is to develop a graph embedding algorithm to protect the edge privacy for non-attributed graphs that contain only node IDs and edges.
\begin{acks}
    This work was partially supported by (i) the National Key Research and Development Program of China 2021YFB3301301,(ii) NSFCGrants U2241211and 62072034. Rong-Hua Li is the corresponding author of this paper.
\end{acks}

\bibliographystyle{ACM-Reference-Format}
\balance
\bibliography{references}


\end{document}